\documentclass[10pt,twocolumn]{article}
\usepackage{graphicx}
\usepackage{geometry} 
\usepackage{color}
\usepackage[square, comma, sort&compress]{natbib}
\pdfoutput=1
\title{An evolutionary model of long tailed distributions in the social sciences}
\author{R. Alexander Bentley\\Anthropology Dept, Durham University\\Durham DH1 3HN UK\\r.a.bentley@durham.ac.uk \and Paul Ormerod\\Volterra Consulting Ltd.\\ London SW14 8AE, UK\\pormerod@volterra.co.uk \and Michael Batty\\Centre for Advanced Spatial Analysis\\ University College London, London, WC1E 6BT, UK\\ m.batty@ucl.ac.uk}
\begin{document}
\maketitle
\textbf{Studies of collective human behavior in the social sciences, often grounded in details of actions by individuals, have much to offer `social' models from the physical sciences concerning elegant statistical regularities. Drawing on behavioral studies of social influence, we present a parsimonious, stochastic model, which generates an entire family of real-world right-skew socio-economic distributions, including exponential, winner-take-all, power law tails of varying exponents and power laws across the whole data.  The widely used Albert-Barab\'{a}si model of preferential attachment is simply a special case of this much more general model. In addition, the model produces the continuous turnover observed empirically within those distributions. Previous preferential attachment models have generated specific distributions with turnover using arbitrary add-on rules, but turnover is an inherent feature of our model.  The model also replicates an intriguing new  relationship, observed across a range of empirical studies, between the power law exponent and the proportion of data represented.}

Since Pareto, the right-skew nature of income distribution has been known, while similar skewness in the frequencies of words, scientific papers, and city sizes have been recognised for decades \cite{1,2,3,4,5}. In the statistical sciences, particularly statistical physics, a recent explosion of interest in such distributions for social phenomena includes internet links \cite{6,7}, author citations \cite{8}, sexual partners \cite{9}, and firm sizes and their extinctions \cite{10,11} amongst many others. 

With socio-economic phenomena, the detailed debate over the exact form of these distributions -- for example, power laws versus similar fat-tailed functions such as the stretched exponential \cite{1,2,13} -- often involves the characterisation of the distribution at a point in time, and often neglects the importance of dynamics and the underlying behaviour \cite{12,14} which gives rise to changes over time within any given distribution.

Simon \cite{3} argued that right-skew distributions were so widespread that their key similarity was likely to be `in the underlying probability mechanisms' that led to their generation. This is clearly the case but, as noted in the social sciences for over a century \cite{12}, it is inherently a description of macro phenomena, without an explanation for the individual behaviour that gives rise to emergent properties. Also, with socio-economic phenomena, the discussion over the exact form of these distributions -- true power laws versus similar fat-tailed functions \cite{1,2,13} -- often neglects the importance of dynamics and their underlying behaviour \cite{12,14}.

We thus propose a model based upon individual agents who are boundedly rational and are influenced by the behaviour of other agents in terms of their decision-making. In other words, the agents act with social purpose, which is fundamentally different from physical or biological phenomena where the agents (or particles) are incapable of intent. The model provides four advances on previous models:

(a) It can generate a wide range of the right-skew distributions observed in cultural, economic and social situations from different combinations of its two parameters.

(b) The widely used Albert-Barab\'{a}si (B-A) model \cite{7} of preferential attachment is simply a special case of this much more general model.

(c) In terms of power law fits, there are two essential statistics, the exponent $\alpha$ and the fraction $f$ of the total observations over which the power law is believed to hold. The model can replicate both observed exponents $\alpha$ and the fraction $f$ from real-world observations \cite{1, 2}. 

(d) Many real-world right-skew distributions exhibit constant turnover in the rankings of their constituents even if their functional form is time-invariant \cite{14, 15}. Unlike the B-A model \cite{7}, our model is capable of generating such turnover without recourse to self-fulfilling rules such as `aging' or variable `fitness' of the individual elements \cite{16}.

\section{The social influence model}
Consider a model populated initially by $N$ agents located in some space such as the sequence of real numbers. Depending on the phenomenon, each location is an abstract representation; it could refer to the city where a firm chooses to locate itself, but it could equally well refer to the product a consumer chooses, or the idea or fashion that a person follows.

The model proceeds in a series of steps. In each step, $n$ new agents enter the model, where the number $n$ is between 1 and $N$ and fixed as a parameter in each solution of the model. With probability $1-\mu$, an agent copies the choice of location from that of an existing agent within the previous m time steps, or else with probability $\mu$, the agent innovates by choosing a unique new location at random. In other words, the agent either copies an existing agent from the last $m$ steps, or chooses a new location. 

Here we restrict our exploration to two key parameters of the model, $m$ and $\mu$, by choosing convenient values for $N$ and $n$. The `memory' parameter $m$ determines the number of steps of the previous decisions of other agents over which an agent looks when making its decision. The `innovation' parameter $\mu$ determines the fraction of the agents who decide to take a completely new decision rather than replicating one of the decisions made by other agents.

\section{Variety of distributions}
A specific version of the model, with $m = 1$ (i.e., memory only of the immediately preceding step), is known in population genetics and physics \cite{17,18}. For the special case of $n = N$ and $m = 1$, analytical solutions demonstrate a power-law distribution \cite{17} for $N\mu$ equal to or slightly greater than 1. For $m = 1$ and $N\mu \ll 1$, this gradually converges on a winner-take-all distribution as $N\mu$ approaches zero.  

The case where $m$ =  all is a further special category of the model, where extinction or obsolescence does not  occur.  In this case, we can achieve different power law slopes by varying $n$ and $\mu$. Figure 1 shows, for example, that we can match the B-A preferential attachment model \cite{7}, obtaining a power law exponent $\alpha \sim 3$ over the entire distribution, by using $m = all$ with $N = 1, n = 10, t = 20,000$, and $\mu = 0.6$.

For socio-cultural phenomena, however, we expect memory to be limited, and thus $m$ in general to take values below the special case of `all'.  So while we define the model to allow $m$ to take any value between 1 and \textit{all}, we explore here a limited range, from $m = 1$ to $m = 100$ time steps of limited memory. The combined effect of varying $m$ along with varying the innovation parameter $\mu$ generates both a wide range of right-skew distributional forms and turnover of rankings of locations within those distributions. Considerable anthropological and socio-economic evidence exists  \cite{19,20,21,22,23} on the plausible values for $\mu$ being no greater than 0.1. 
  
  \begin{figure}
  \begin{center}
  \includegraphics[width=3in]{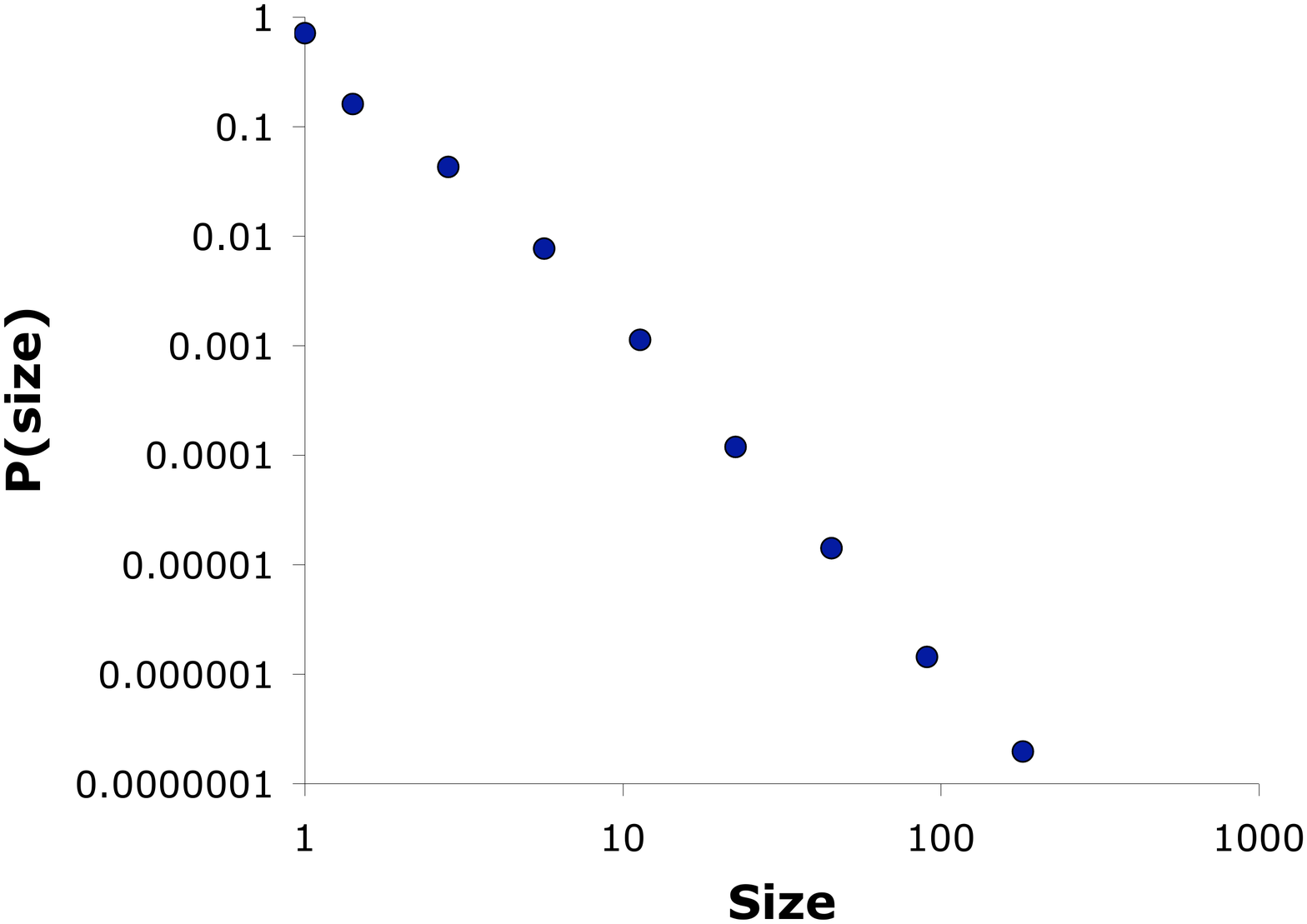}
  \end{center}
  \caption{The power law generated from the preferential attachment version of the  model. As the probability distribution for a typical model run using $N = 1, n = 10,   t = 20,000, \mu = 0.6$, and $m =$ all (where the generated sizes are logarithmically binned). The exponent for the power law is Ð2.9 ($r^2 = 0.996$), matching that reported (also by least-squares regression) for preferential attachment models \cite{7}.}
  \end{figure}

Figure 2a plots typical solutions of the model using acceptable values of $\mu$, while varying $m$ (holding $N = 1000$, $n = 100$ and showing the results at time step 1000). Aside from the selected results shown in this figure, the model produces additional results ranging from a winner-take-all outcome, to a power law over the entire distribution (exponent $\alpha \sim 1.5$), to a power law fitted to the tail of varying exponent. Figure 2b illustrates how the model parameters can be selected so that the results match real-world right-skew distributions, such as religions, website subscriptions, word use, names, and author citations.
    
\begin{figure}
  \begin{center}
\includegraphics[width=3in]{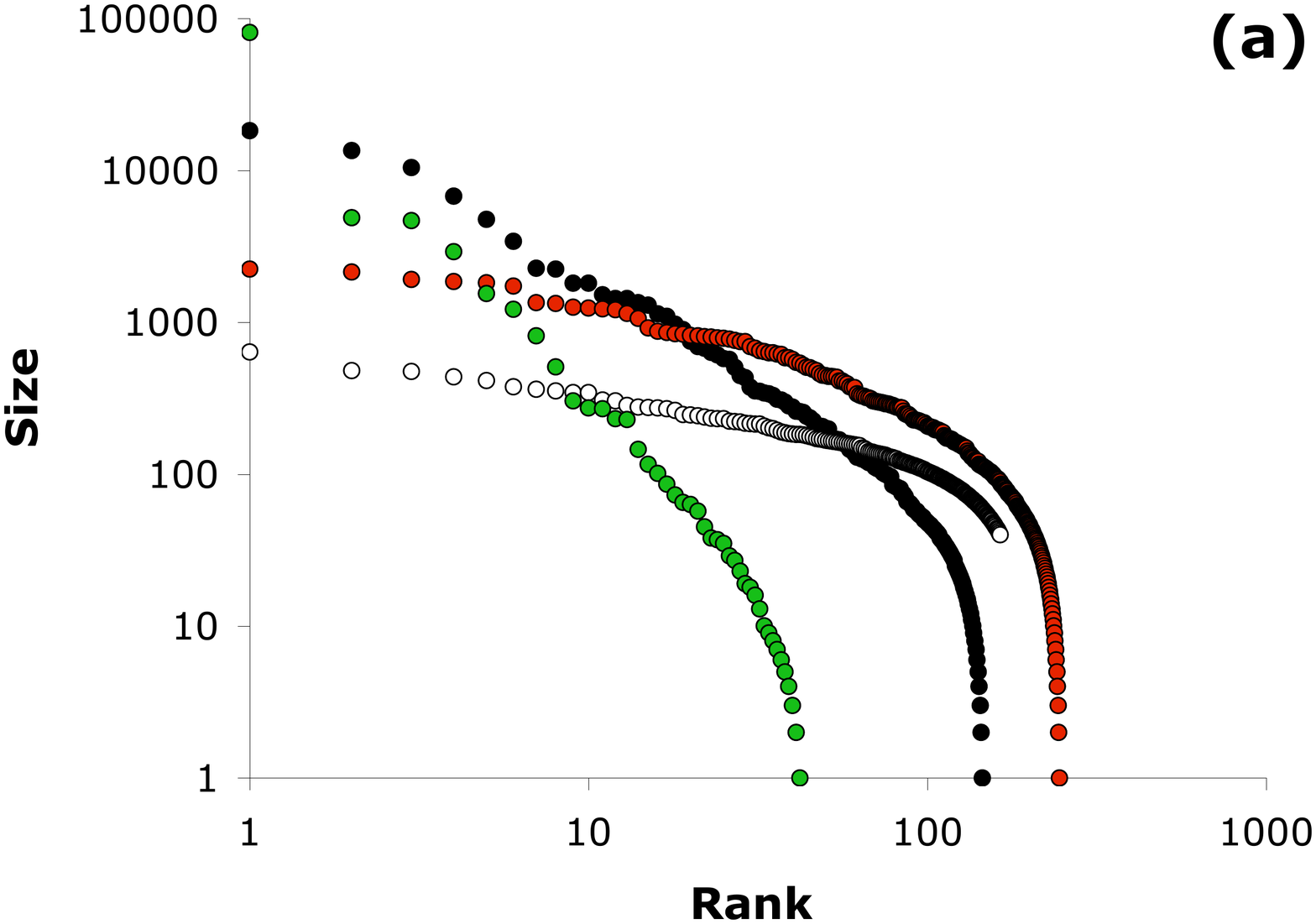}
\includegraphics[width=3in]{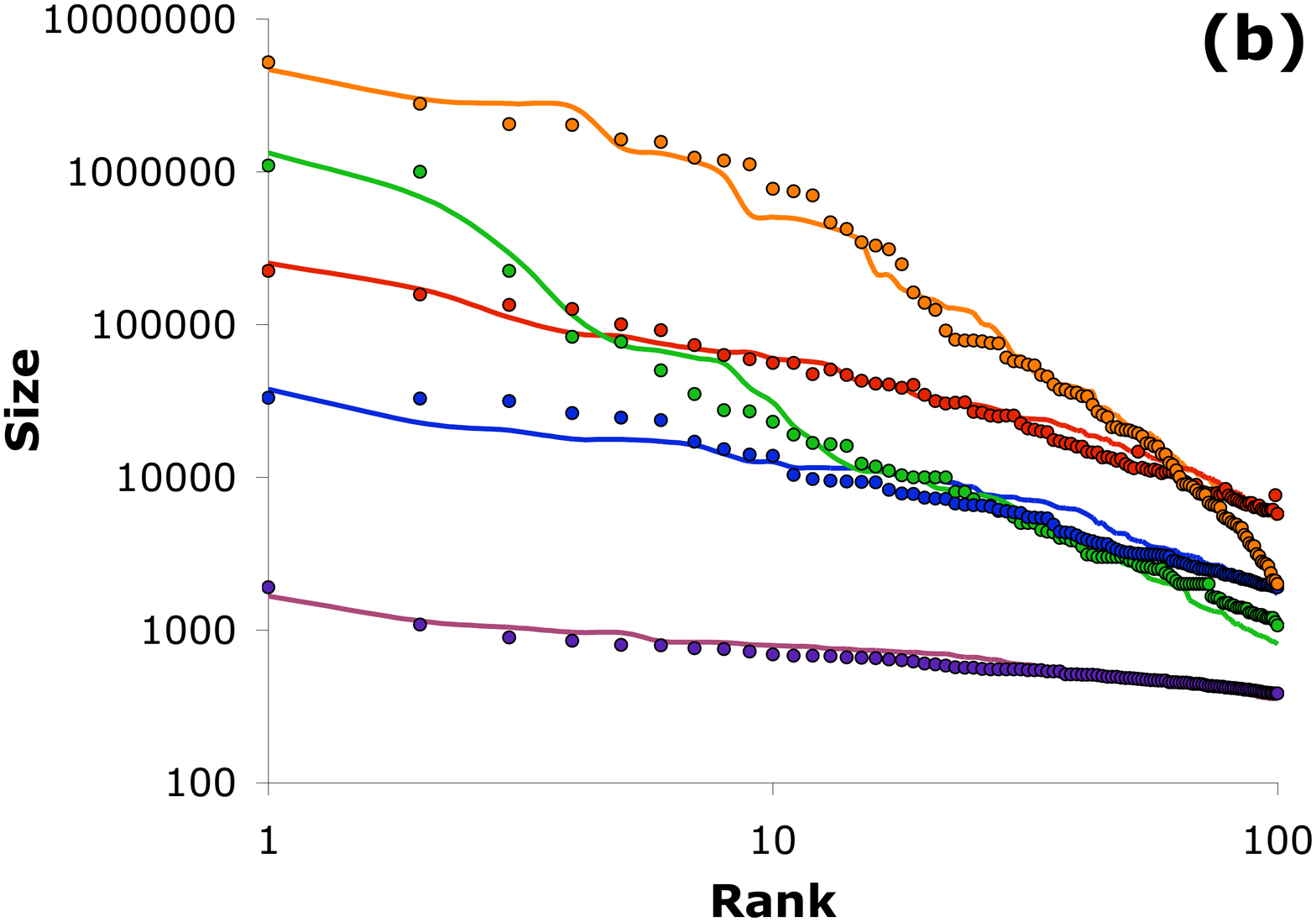}
  \end{center}
 \caption{Log-log plots of rank and size. \textbf{(a)} for typical model solutions with $N = 1000, n = 100, t = 1000$ and: $\mu = 0.01, m = 1$ (black); $\mu = 0.01, m = 100$ (red); $\mu = 0.08, m = 100$ (white); $\mu = 0.0001, m = 2$ (green). \textbf{(b)} for real-world top 100 ranked lists (dots) versus model results (lines). Top 100 lists include \cite{top100}: male baby name frequency (per million) in the 1990 US census (blue), RSS feed subscriptions 2001-2008 (orange), English words (red), cited economists 1993-2003 (purple), and religions in thousands of adherents (green). With $N = 1000$, the model fits were made with $\mu = 0.001, m = 50, n = 200, t = 4000$ for names, $\mu = 0.00002, m = 6, n = 2500, t = 10000$ for RSS feeds, $\mu = 0.00025, m = 85, n = 100, t = 1100$ for cited economists, $\mu = 0.004, m = 4, n = 450, t = 8000$ for words, and $\mu = 0.0007, m = 2, n = 100, t = 4000$ for religions.}
 \end{figure}
 
\section{Regularity in the long tail}
Table 1 lists power law tail exponents $\alpha$ for various recently collated social data sets \cite{1,2} along with the fraction $f $($= n_{tail}/n$) of total observations in the tail. A striking, and previously unreported, feature of these estimates is the relationship between $\alpha$ and $f$, where these data reveal a clear inverse correlation. The smaller the fraction $f$ of the distribution best-fit to a power law tail \cite{24}, the larger the exponent $\alpha$ of that tail. The least-squares fit is $\alpha  \cong 1.54f^{-0.156}$ ($r^2 = 0.952$). 

Figure 3 plots this relationship in the empirical data along with the least squares fit using the model, as solved 100 times, for each of $\mu =$ 0.05, 0.06 and 0.07, with $m = 30$ in each case (and $N = 1000$, $n = 100$, $t = 1,000$). The results show $\alpha \cong 1.56f^{-0.155}$ ($r^2 = 0.975$), very similar to the data-based relationship.

\begin{figure}
 \begin{center}
 \includegraphics[width=3in]{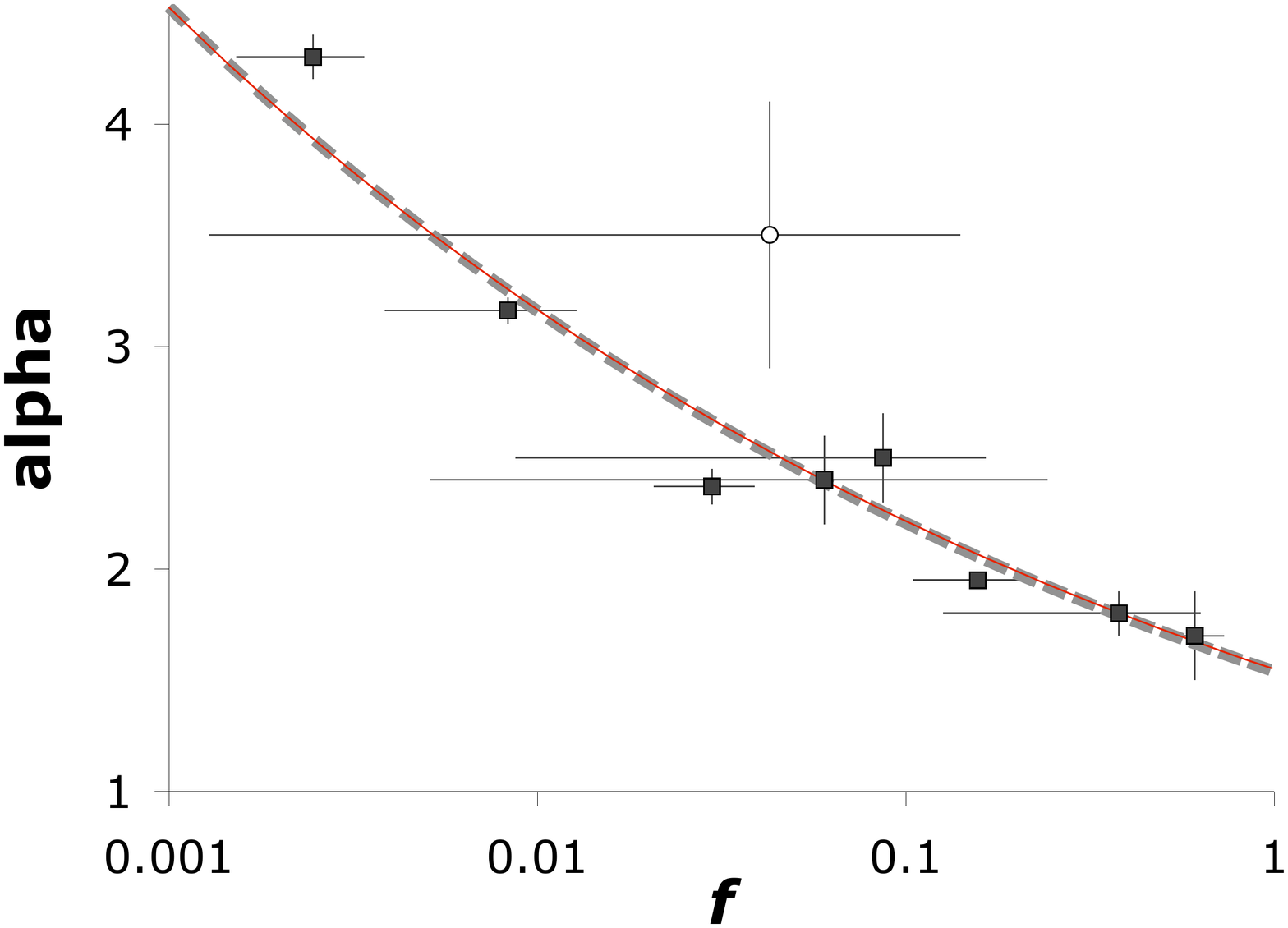}
 \end{center}
 \caption{The power law tail exponent $\alpha$ versus the fraction $f$ of total observations represented by the tail. The dots show power law tails calculated for various real-world socio-cultural data sets (see Table 1 for values and errors), whose relationship (dashed grey curve) can be approximated by $\alpha = 1.54f^{-0.156}$ ($r^2 = 0.952$ except for the outlier -- the open circle -- from email lists). The thin red curve shows the least squares fit from 300 runs of our theoretical model which gives $\alpha = 1.56/f^{-0.155}$ ($r^2 = 0.975$). Exponents have been estimated using maximum likelihood \cite{1,2}.}
  \end{figure}

\section{Distribution of turnover} 
The model also produces continual turnover through time for any given distribution as demonstrated by the distributions of lifespans within ranked lists as in Figure 4a. This resembles the lifespans of real world social and economic fat-tail distributions in Figure 4b. The memory parameter $m$ again expands the power of the model. Although turnover has already been demonstrated \cite{17} for the special case $m = 1$, different values of m are needed to account for empirically observed turnover (Supplementary Information shows distributions generated for increasing memory $m$ with order of magnitude changes in the value of $\mu$).

\begin{table*}
\caption{\bf Power-law fits determined by \cite{2} among socio-cultural data sets}Parameters include number of observations $n$, maximum observed value $x_{max}$, observations in the tail $n_{tail}$ and the minimum value in the tail $x_{min}$
\begin{tabular}{@{\vrule height 8pt depth5pt  width0pt}lrccccc}
\\\hline
\textbf{Quantity} & $n$ & $x_{max}$ & $x_{min}$ & $\alpha$ & $n_{tail}$ & $f = n_{tail}/n$\\\hline
intensity of wars &115 & 382 & $2.1 \pm 3.5$ & $1.7 \pm 0.2$ & $70 \pm 14$ & 0.609\\
religious followers (x $10^6$)	&103 & 1050 & $3.85 \pm 1.60$ & $1.8 \pm 0.1$ & $39 \pm 26$ & 0.379\\
word count & 18855 & 14086 & $7 \pm 2$ & $1.95 \pm 0.02$ & $2958 \pm 987$ & 0.157\\
city population (x $10^3$) & 19447 & 8009 & $52.5 \pm 11.9$ & $2.37 \pm 0.08$ & $580 \pm 177$ & 0.030\\
terrorist attack severity & 9101 & 2749 & $12 \pm 4$ & $2.4 \pm 0.2$ & $547 \pm 1663$ & 0.060\\
surname frequency (x $10^3$) & 2753 & 2502 & $112 \pm 41$ & $2.5 \pm 0.2$ & $239 \pm 215$ & 0.087\\
paper citations & 415229 & 8904 & $160 \pm 35$ & $3.16 \pm 0.06$ & $3455 \pm 1859$ & 0.008\\
email address books & 4581 & 333 & $57 \pm 21$ & $3.5 \pm 0.6$ & $196 \pm 449$ & 0.043\\
papers authored & 401455 & 1416 & $133 \pm 13$ & $4.3 \pm 0.1$ & $988 \pm 377$ & 0.002\\
\end{tabular}
\end{table*}

\begin{figure}
 \begin{center}
  \includegraphics[width=2.7in]{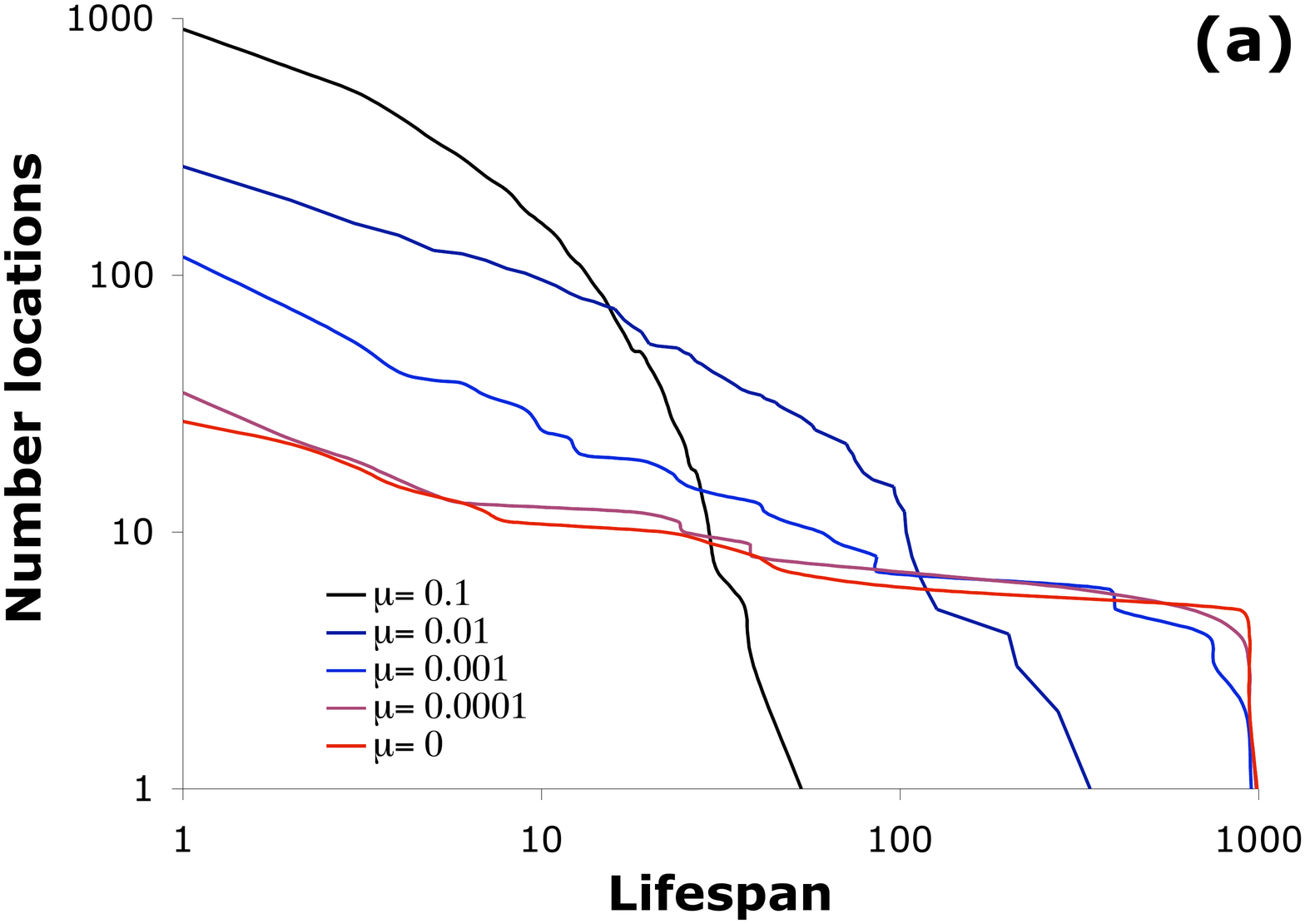}
\includegraphics[width=2.7in]{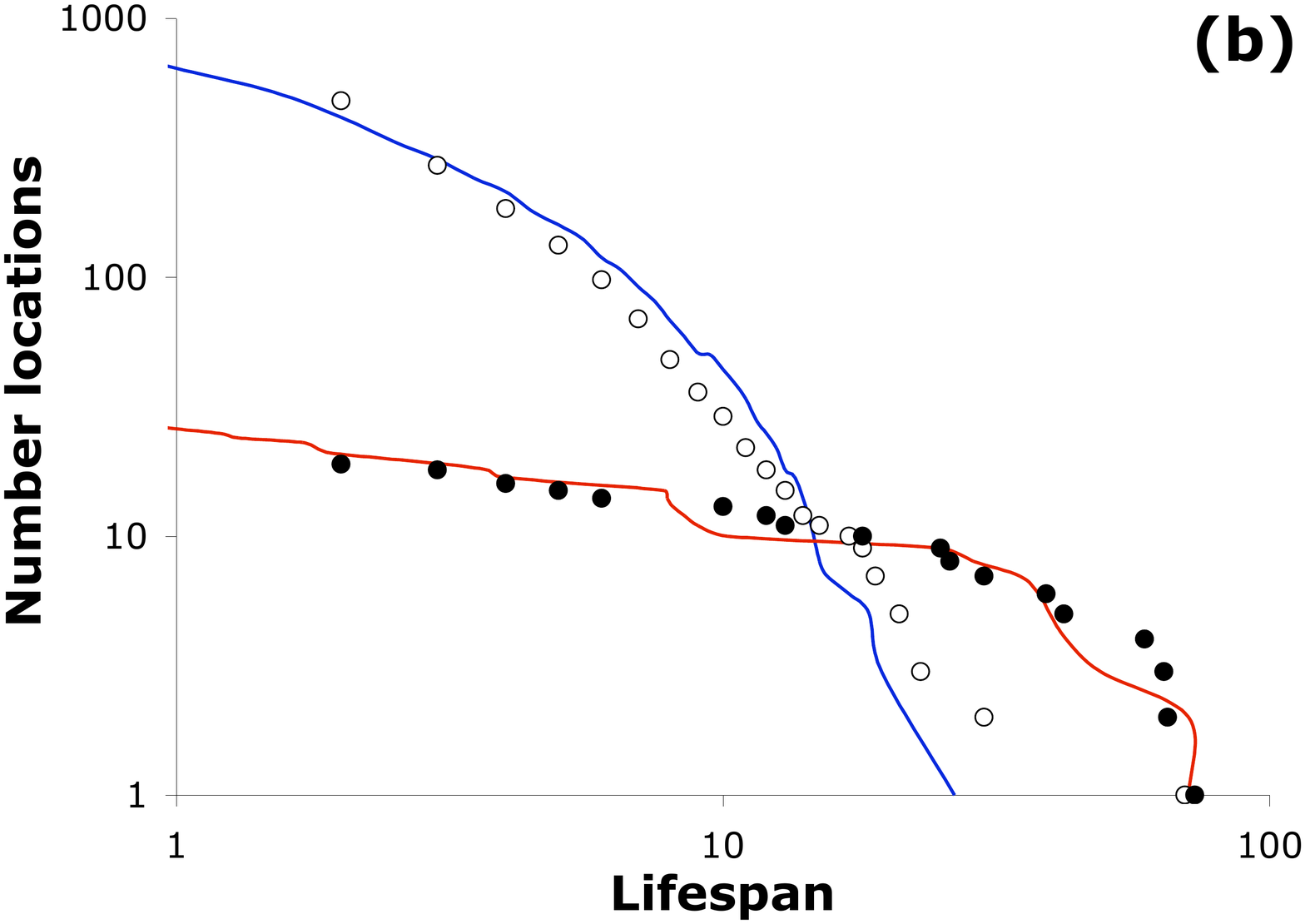}
 \end{center}
 \caption{Life-spans of individual locations. \textbf{(a)} Typical model runs, showing the cumulative distribution of number of time steps spent in the top 5 for model runs of 1000 time steps with $N = 1000, n = 100$, and $m = 1$. \textbf{(b)} Life-spans of UK Number One Hits \cite{UK1s} for 1956-2007 (open circles), versus the model, $m = 1, \mu = 0.1$ (blue line), and $t$ years in the Top 5 US boys' names \cite{babynames}, 1907-2006 (filled circles) versus the model, $m = 10, \mu = 0.001$ (red line). Since the temporal units are arbitrary, the modelled lifespans were divided by 2 to match the albums, and divided by 10 to match the names (which on the log-log plot slides the distribution to the left).}
  \end{figure}
  
\section{Discussion}
The model we have presented can generate not only a wide range of long-tailed distributions but a constant turnover of the constituent agents within any given overall rank-size distribution. It is also able to replicate a newly-identified empirical relationship whereby the power law exponent increases as the proportion of data in the tail falls. 

The model is quite general, despite using only two parameters. Varying the parameter values can yield a range of distributions, such as a power law over the whole sample, a power law only in the tail, and a winner-take-all outcome. Since the parameter $m$ represents memory and $\mu$ represents innovation in modelled decision-making, the real-world relationship between $\alpha$ and $f$ in Figure 3 may result from a variation in related parameters among the different contexts of human decision-making. We conjecture that the continuous relationship observed in Figure 3 suggests that socio-economic power law distributions may form a continuum resulting from a generalised process with limited memory. In contrast to the special $m = all$ case (Figure 1), when model runs with limited memory yield a power law over the entire distribution ($f = 1$), it is only with exponent $\alpha$ close to 1.5 (Figure 3).

This combination of results makes this model unique among the many alternatives that can produce power laws. The most commonly proposed processes such as preferential attachment, proportionate effect based on Gibrat's principle, the `Matthew effect' and the Yule process \cite{1,2,14,25,26}, produce power laws from the positive feedback introduced by interactions between individual agents. But these `rich get richer' models have not been able to account for flux in the constituents of the ranked distribution \cite{27}, either when growth is one of strict preferential attachment or even when growth is proportionate to a stochastic rate independent of size \cite{28}. Even though ``dynamical problems lie at the forefront" of network science \cite{16}, in most network models, existing connections affect future connections such that change does not occur naturally, but only with imposed modifications. 

Social scientists have been critical of modelling social and economic data by mapping onto known phenomena in physics without considering realistic behavioural motivations of the agents \cite{12,29,30}. As a step in this direction, our model captures two fundamental motivations, the imitation of others and novelty in invention. 

Compared to similar, less flexible versions of this model \cite{15}, a crucial new variable appears to be the memory $m$, which reflects different time frames to which agents will refer in different contexts. In terms of pure fashion markets such as popular music for example \cite{Salganik2006}, agents take into account only the most recent decisions of others and hardly ever those of several months or even weeks ago. However in choosing where to locate geographically, for example, a firm or a person in a city will implicitly be using information from many previous time steps with respect to the decisions made by others.

Generating a range of long-tailed distributions with dynamic turnover, these features distinguish this model from the standard socio-economic science model of individual rational behaviour where social influence is the exception to the rule (as in, for example, `irrational' stock market bubbles or real estate crises). With its unrealistic psychological assumptions \cite{Kahneman2003} and inconsistencies with experimental results \cite{Smith2003}, the standard model suffers from a neglect of social influence, even in its modern form which permits, for example, asymmetry in the amount of information possessed by different agents \cite{Akerlof1970,Stiglitz2002}, the cost of gathering information \cite{Stigler1961}, and imperfections in gathering and processing information \cite{3}. 

Social influence is arguably ubiquitous among the human species \cite{DunbarShultz2007}. In fact, rather than the agent's cost-benefit analysis that has served as a null hypothesis for rationality for over a century, an alternative is that each agent uses (consciously or not) the decisions of others as a basis for his or her own decisions. 

The social-influence model we have presented allows choices among multiple possible alternatives, which rise and fall in relative popularity over time, rather than binary, `either-or' decisions. This is truly reflective of human interactions such as the choice of a popular name for a child, the citation of an academic paper, or movement to a city where others have chosen to live. Indeed, these phenomena are inherently defined by the past decisions of others, without which there would be no cities, familiar names, or popular culture.

\section{Acknowledgments}
Batty was partially supported by the EPSRC Spatially Embedded Complex Systems Engineering Consortium (EP/C513703/1). Amy Heineike of Volterra provided programming support.

\end{document}